\documentclass[Physsubmission, Phys]{SciPost}

\usepackage{xcolor}
\colorlet{alert}{red!60!black}
\colorlet{example}{green!60!black}
\colorlet{structure}{blue!60!black}

\newcommand{\ind}[1]{\rm\scriptscriptstyle #1}

\def\lsim{\mathrel{\rlap{\lower4pt\hbox{\hskip1pt$\sim$}}
    \raise1pt\hbox{$<$}}}                
\def\gsim{\mathrel{\rlap{\lower4pt\hbox{\hskip1pt$\sim$}}
    \raise1pt\hbox{$>$}}}                

\newcommand{\mnote}[1]{}                   

\hypersetup{
    colorlinks,
    linkcolor={red!50!black},
    citecolor={blue!50!black},
    urlcolor={blue!80!black}
}

\usepackage[bitstream-charter]{mathdesign}
\DeclareSymbolFont{usualmathcal}{OMS}{cmsy}{m}{n}
\DeclareSymbolFontAlphabet{\mathcal}{usualmathcal}

\begin{document}

\begin{center}{\Large \textbf{The Compton Amplitude,
               lattice QCD and the Feynman--Hellmann approach \\
}}\end{center}

\begin{center}
K.~U. Can\textsuperscript{1},
A. Hannaford-Gunn\textsuperscript{1},
R. Horsley\textsuperscript{2$\star$},
Y. Nakamura\textsuperscript{3},
H. Perlt\textsuperscript{4},
P.~E.~L. Rakow\textsuperscript{5}, \\
E. Sankey\textsuperscript{1},
G. Schierholz\textsuperscript{6},
H. St\"uben\textsuperscript{7},
R.~D. Young\textsuperscript{1} and
J.~M. Zanotti\textsuperscript{1}
\end{center}

\begin{center}

{\bf 1} CSSM, Department of Physics,
        University of Adelaide, Adelaide SA 5005, Australia \\
{\bf 2} School of Physics and Astronomy, University of Edinburgh,
        Edinburgh  EH9 3FD, UK \\
{\bf 3} RIKEN Center for Computational Science,
        Kobe, Hyogo 650-0047, Japan \\
{\bf 4} Institut f\"ur Theoretische Physik,
        Universit\"at Leipzig, 04109 Leipzig, Germany \\
{\bf 5} Theoretical Physics Division,
        Department of Mathematical Sciences, \\
        University of Liverpool,
        Liverpool L69 3BX, UK \\
{\bf 6} Deutsches Elektronen-Synchrotron DESY,
        Notkestr. 85, 22607 Hamburg, Germany \\
{\bf 7} Universit\"at Hamburg, Regionales Rechenzentrum,
        20146 Hamburg, Germany \\
* rhorsley@ph.ed.ac.uk
\end{center}

\begin{center}
\today
\end{center}


\definecolor{palegray}{gray}{0.95}
\begin{center}
\colorbox{palegray}{
  \begin{minipage}{0.95\textwidth}
    \begin{center}
    {\it  XXXIII International (ONLINE) Workshop on High Energy Physics \\
  “Hard Problems of Hadron Physics:  Non-Perturbative QCD \& Related Quests”}\\
    {\it November 8-12, 2021} \\
    \doi{10.21468/SciPostPhysProc.?}\\
    \end{center}
  \end{minipage}
}
\end{center}

\section*{Abstract}
{\bf
A major objective of lattice QCD is the computation of hadronic
matrix elements. The standard method is to use three-point and
four-point correlation functions. An alternative approach, 
requiring only the computation of two-point correlation functions 
is to use the Feynman-Hellmann theorem. In this talk we develop this 
method up to second order in perturbation theory, in a context appropriate 
for lattice QCD. This encompasses the Compton Amplitude (which forms the
basis for deep inelastic scattering) and hadron scattering.
Some numerical results are presented showing results indicating what 
this approach might achieve.
}



\section{Introduction}


Understanding the internal structure of hadrons and in particular
the nucleon directly from the underlying QCD theory is a major task
of particle physics. It is complicated because of the non-perturbative
nature of the problem, and presently the only known method is to discretise
QCD and use numerical Monte Carlo methods. The relevant information is
encoded in correlation functions -- from the all encompassing 
two-quark correlation functions to GTMDs, TMDs, GPDs Wigner functions,
PDFs and Form Factors, e.g.\ \cite{Diehl:2015uka}.

Using the Operator Product Expansion (OPE), it is possible to relate
form factors to moments of certain matrix elements, which in principle
are calculable using lattice QCD techniques. However due to theoretical
problems such as much more mixing of lattice operators due to reduced $H(4)$ 
symmetry and numerical problems, for many years it was only possible 
to compute the very lowest moments. (As an example of a complete 
calculation -- albeit for quenched fermions -- see for example 
\cite{Gockeler:2004wp}.) This does not allow for the reconstruction 
of the associated PDF. Progress was recently achieved with the concepts 
of quasi-PDFs and pseudo-PDFs, for a comprehensive review 
see \cite{Cichy:2018mum}.

Here in this talk we shall describe a complementary
approach which relates the structure function to that of the
associated Compton amplitude, emphasising via dispersion relations the physical
and unphysical regions and their connection with Minkowski and Euclidean
variables. While the Compton amplitude is a correlation function 
it is $4$-point and hence difficult to compute with the straightforward 
standard approach used in Lattice QCD of tying the appropriate
Grassmann quark lines together in the path integral. However,
we are able to circumvent this problem by using a Feynman--Hellmann 
approach. This approach avoids operator mixing problems, has a simple
renormalisation and as independent of the Operator Product
Expansion, OPE, allows an investigation of power corrections to the
leading behaviour (twist $2$) of the OPE. We first described this 
method in \cite{Chambers:2017dov} and have been developing it further 
e.g.\ \cite{Can:2020sxc,Can:2021tgi}. 

In this talk we give a brief introduction to this approach, first in
section~\ref{SF+CA} giving the relation between structure functions
and the Compton amplitude. This is followed in section~\ref{FH}
by a description of the Feynman--Hellmann approach. Some numerical
results are given in \ref{results}. Further details and results are given
in \cite{Can:2020sxc}. The Feynman--Hellmann approach is a versatile method
and in the following section~\ref{further_applications} some further 
applications are mentioned. Finally we give some conclusions.


\section{Structure functions and the Compton amplitude}
\label{SF+CA}

\begin{figure}[htb]

   \begin{center}
      \includegraphics[width=5.00cm]{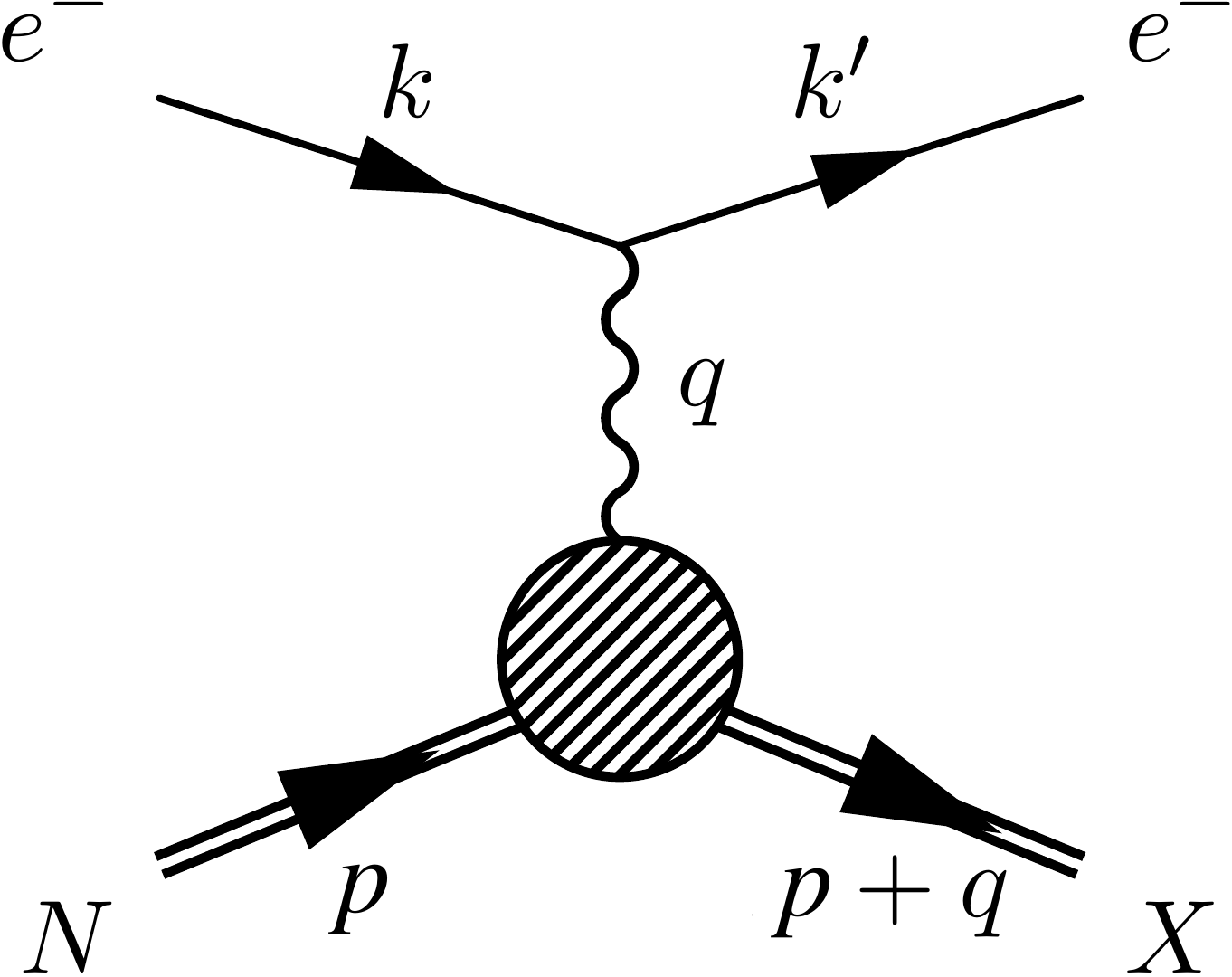}
   \end{center} 

\caption{DIS, where $k$, $k^\prime$ represent the incoming, outgoing
         lepton momenta, $p$ is the momentum of the incoming nucleon
         of mass $M_N$, $q = k - k^\prime$ is the momentum transfer and 
         $X$ represents the recoiling system.}
\label{dis+contour}
\end{figure}
\begin{figure}[htb]
\begin{minipage}{0.45\textwidth}

   \begin{center}
      \includegraphics[width=5.00cm]{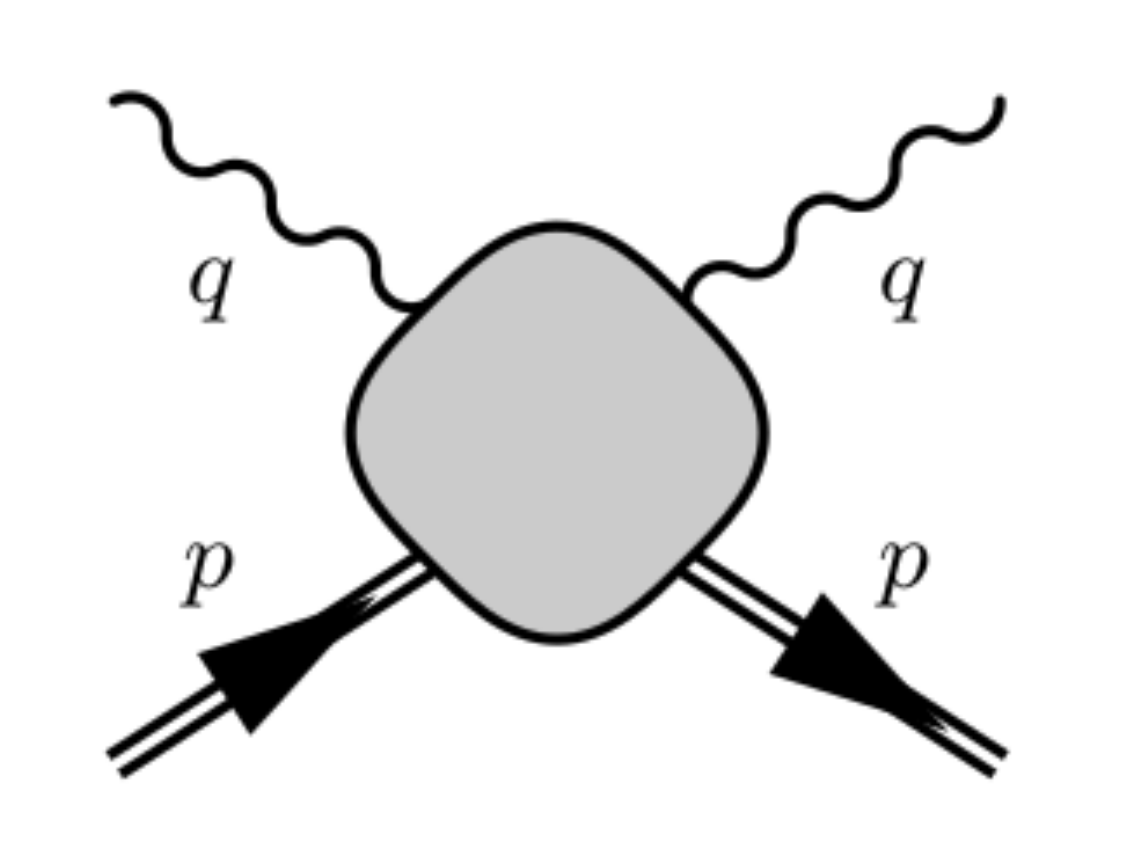}
   \end{center} 

\end{minipage}\hspace*{0.10\textwidth}
\begin{minipage}{0.50\textwidth}

   \begin{center}
      \hspace*{-1.50in}
      \includegraphics[width=6.00cm]{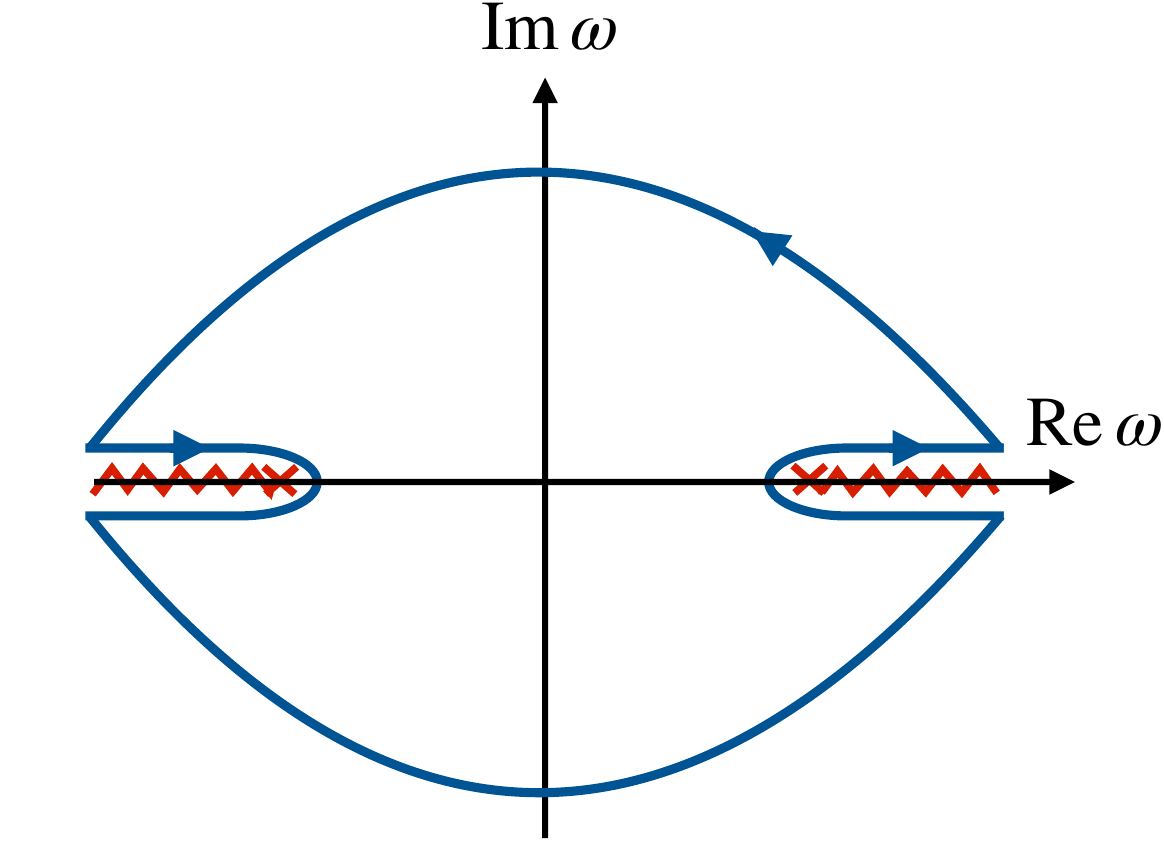}
   \end{center} 

\end{minipage}
\caption{LH panel: The forward Compton Amplitude. 
         RH panel: The analytic structure for ${\cal F}_1$ 
                   -- branch cuts starting from $\omega = \pm1$, together
                   with the contour used for the dispersion relation.}
\label{compton+contour}
\end{figure}

Deep Inelastic Scattering (DIS) is the inclusive scattering of a lepton
(usually an electron) from  nucleon (usually a proton), $eN \to e^\prime X$.
The process is shown diagrammatically in Fig.~\ref{dis+contour}.
The kinematics is such that $Q^2 \equiv -q^2 > 0$; the invariant mass of
$X$ is $M_X^2 = (p+q)^2$ and the Bjorken variable, $x$, is defined by
$x = Q^2/(2p\cdot q)$. Here we shall be mainly using the inverse Bjorken 
variable $\omega = 1/x$. $x > 0$ from kinematics and $M_X ^2 > M_N^2$ means
that $x < 1$ which translates to $1 < \omega < \infty$ as the
physical region. The square of the amplitude can be factorised into
a calculable leptonic tensor together with an unknown hadronic tensor, 
$W_{\mu\nu}$, given%
\footnote{The state normalisation is given by
          ${}_{\rm rel}\langle N|N \rangle_{\rm rel} = 2E_N$. 
          See also footnote~\ref{norm_conv}.}
by
\begin{eqnarray}
   W_{\mu\nu} 
     \equiv {1 \over 4\pi} \int d^4z \, e^{iq\cdot z} \rho_{ss^\prime} \,
            {}_{\rm rel}\langle p, s^\prime|[J_\mu^\dagger(z), J_\nu(0)]
                                        |p, s\rangle_{\rm rel} \,,
\end{eqnarray}
where $J_\mu$ is the electromagnetic current ($\gamma$)%
\footnote{This can, of course, be generalised to neutral ($Z$)
or charged ($W^{\pm}$) currents.}
and for unpolarised nucleons we have $\rho_{ss^\prime} = \delta_{ss^\prime}/2$. 
The tensor has the Lorentz decomposition
\begin{eqnarray}
   W_{\mu\nu} =
   \left(-\eta_{\mu\nu} + {q_\mu q_\nu \over q^2} \right) F_1(x,Q^2)
          + \left(p_\mu - {p\cdot q \over q^2}q_\mu \right)
             \left(p_\nu - {p\cdot q \over q^2}q_\nu \right)
              { F_2(x,Q^2) \over p\cdot q} \,,
\end{eqnarray}
with structure functions $F_1(x,Q^2)$ and $F_2(x,Q^2)$.
It is useful to relate the $W_{\mu\nu}$ scattering amplitude to the
forward Compton scattering amplitude, $T_{\mu\nu}$, depicted in the LH panel
of Fig.~\ref{compton+contour}, as this is a correlation function
and so more amenable to lattice QCD or other calculational methods.
The definition parallels that of $W_{\mu\nu}$
\begin{eqnarray}
   \lefteqn{T_{\mu\nu}(p,q) 
      \equiv i \int d^4z \, e^{iq\cdot z} \rho_{ss^\prime} \,
           {}_{\rm rel}\langle p, s^\prime
                        |T(J_\mu^\dagger(z) J_\nu(0))|p, s\rangle_{\rm rel} }
      & &                                               \nonumber  \\
      &=& \left(-\eta_{\mu\nu} + {q_\mu q_\nu \over q^2} \right) 
                  {\cal F}_1(\omega,Q^2)
          + \left(p_\mu - {p\cdot q \over q^2}q_\mu \right)
             \left(p_\nu - {p\cdot q \over q^2}q_\nu \right)
              { {\cal F}_2(\omega,Q^2) \over p\cdot q } \,,
\label{T_mink}
\end{eqnarray}
with corresponding structure functions ${\cal F}_1(\omega,Q^2)$,
${\cal F}_2(\omega,Q^2)$. Due to the time ordering in its definition
it is a correlation function. These are related via the Optical theorem
to the hadronic tensor structure functions by 
${\rm Im} {\cal F}_1(\omega, Q^2) = 2\pi F_1(x,Q^2)$ (and similarly
for $F_2$, however in this talk we shall concentrate on $F_1$).
Photon crossing symmetry $N \to \bar{N}$ means that ${\cal F}_1$ is
symmetric under $\omega \to - \omega$ (while ${\cal F}_2$ is
anti-symmetric). The analytic structure, e.g. \cite{gasiorwicz},
is thus given in the RH plot of 
Fig.~\ref{compton+contour}. Analyticity properties (including using 
the Schwarz reflection principle across the branch cut) then give 
a once subtracted%
\footnote{Conventionally $\omega = 0$ is chosen as the subtraction
point, but others have recently been suggested, \cite{Hagelstein:2020awq}.}
dispersion relation
\begin{eqnarray}
   {\cal F}_1(\omega,Q^2)
      &=& {2\omega \over \pi}
          \int_1^\infty d\omega^\prime \left[
             { {\rm Im}{\cal F}_1(\omega^\prime,Q^2) \over
              \omega^\prime(\omega^\prime - \omega-i\epsilon) }
             - { {\rm Im}{\cal F}_1(\omega^\prime,Q^2) \over
              \omega^\prime(\omega^\prime + \omega-i\epsilon) } 
                                    \right] + {\cal F}_1(0,Q^2)
                                                             \nonumber \\
      &=& \underbrace{ 4\omega^2 \int_0^1 dx^\prime 
            { x^\prime F_1(x^\prime,Q^2) \over 
              1 - x^{\prime\,2}\omega^2 - i\epsilon}}_{\overline{\cal F}_1(\omega,Q^2)}
                 + \underbrace{{\cal F}_1(0,Q^2)}_{\rm once\,\,subtracted} \,.
\label{disp_rel}
\end{eqnarray}
(Replacing $x^\prime F_1$ by $F_2$ with no subtraction gives the equivalent
dispersion relation for $F_2$.)
As long as we are in the unphysical region 
$|\omega| < 1  \Longleftrightarrow M_X^2 < M_N^2$,
i.e.\ below elastic threshold, there is no singularity in previous integral
the time ordering is irrelevant, so the $i\epsilon$ in eq.~(\ref{disp_rel})
can be dropped. The Minkowski and Euclidean amplitudes are then identical
which as we shall see in section~(\ref{FH}) will eventually allow a 
direct lattice QCD computation. Physically $|\omega| < 1$ means states 
propagating between currents cannot go on-shell.
Taylor expanding the denominator in eq.~(\ref{disp_rel}) then gives
\begin{eqnarray}
   {\overline{\cal F}}_1(\omega, Q^2) 
      = 2\sum_{n=1}^\infty \omega^{2n} M^{(1)}_{2n}(Q^2)\,,
   \quad \mbox{where} \quad
   M^{(1)}_{2n}(Q^2) = 2\int_0^1 dx^\prime\, x^{\prime\,2n-1} F_1(x^\prime,Q^2)
\label{mellin_moments}
\end{eqnarray}
are the Mellin moments of $F_1$. Furthermore for the numerical results 
considered later we set $\mu = \nu = z$, $p_z = q_z = 0$ giving
\begin{eqnarray}
   T_{33}(p,q) = {\overline{\cal F}}_1(\omega, Q^2)
              = 2\sum_{n=1}^\infty \omega^{2n} M^{(1)}_{2n}(Q^2) \,.
\label{T33}
\end{eqnarray}
So from Compton amplitude data we can directly extract the Mellin moments.
The positivity of the cross section means that $F_1 > 0$ or
$M^{(1)}_{2} \ge M^{(1)}_{4} \ge \ldots M^{(1)}_{2n} \ge \ldots > 0$ so
the expected shape of the Compton amplitude in the unphysical region
for fixed $Q^2$ is simply an increasing polynomial function of $\omega^2$.


\section{The Feynman--Hellmann approach}
\label{FH}


The task now is to compute the (Euclidean) Compton amplitude
and in particular that given in eq.~(\ref{T33}). A direct lattice QCD
computation of the path integral for the necessary 4-point correlation 
function is complicated as there are many diagrams to compute. 
As an alternative we shall use the Feynman--Hellmann approach here.

We now sketch a derivation of the procedure. Consider the $2$-point 
nucleon correlation function
\begin{eqnarray}
   C_{fi\,\lambda}(t;\vec{p},\vec{q}) 
      = {}_\lambda\langle 0| 
           \underbrace{\hat{\tilde{B}}_{N_f}(0;\vec{p})}_{\rm Sink:\, momentum}
                           \hat{S}(\vec{q})^t 
           \underbrace{\hat{\bar{B}}_{N_i}(0,\vec{0})}_{\rm Source:\, spatial} 
                                 |0\rangle_\lambda \,,
\end{eqnarray}
where $\hat{S}$ is the $\vec{q}$-dependent transfer matrix
$\hat{S}(\vec{q}) = \exp{(-\hat{H}(\vec{q}))}$ in the presence of a 
perturbed Hamiltonian
\begin{eqnarray}
   \hat{H}(\vec{q}) 
      = \hat{H}_0 
         - \sum_\alpha \lambda_\alpha \hat{\tilde{{\cal O}}}_\alpha(\vec{q}) \,,
\end{eqnarray}
where
\begin{eqnarray}
   \hat{\tilde{{\cal O}}}_\alpha(\vec{q})
      = \int_{\vec{x}} \left( \hat{O}_\alpha(\vec{x})e^{i\vec{q}\cdot\vec{x}}
                  + \hat{O}_\alpha^\dagger(\vec{x}) e^{-i\vec{q}\cdot\vec{x}}
                     \right)
\end{eqnarray}
is a Hermitian operator. $\lambda$ can be taken as a real positive 
parameter%
\footnote{Can generalise to complex $\lambda$ by absorbing the phase
into the operator:
   $\lambda_\alpha\hat{O}_\alpha(\vec{x})
   \to |\lambda_\alpha|e^{i\phi_{\alpha}}\hat{O}_\alpha(\vec{x})$.}.
Using time dependent perturbation theory via the Dyson Series, namely
the operator expansion, regarding $\hat{B}$ as `small'
\begin{eqnarray}
    e^{t(\hat{A}+\hat{B})}
      = e^{t\hat{A}} 
           + \int_0^t dt^\prime\, e^{(t-t^\prime)\hat{A}} \, 
                    \hat{B} \, e^{t^\prime\hat{A}}
           + \int_0^t dt^\prime\,\int_0^{t^\prime} dt^{\prime\prime} \,
                 e^{(t-t^\prime)\hat{A}} \, \hat{B} \, e^{(t^\prime-t^{\prime\prime})\hat{A}}
                    \, \hat{B}e^{t^{\prime\prime}\hat{A}}
           + O(\hat{B}^3) \,,
\end{eqnarray}
and inserting complete sets of unperturbed states%
\footnote{The lattice normalisation is used here: 
$\langle X(\vec{p}_X)|Y(\vec{p}_Y)\rangle = \delta_{XY}\delta_{\vec{p}_X\vec{p}_Y}$.
To convert to the usual relativistic normalisation, with an additional factor
$2E_X$, change
$|X\rangle \to  |X\rangle / \sqrt{\langle X|X \rangle}$ with
$|0\rangle \to |0\rangle$. \label{norm_conv}}
\begin{eqnarray}
   \sum_{\vec{p}} |N(\vec{p})\rangle\langle N(\vec{p})| 
      + \sum_{E_X(\vec{p}_X) > E_N(\vec{p})} |X(\vec{p}_X)) 
            \rangle \, \langle X(\vec{p}_X)|
                = 1 \,,
\end{eqnarray}
appropriately gives after some algebra the factorised result
\begin{eqnarray}
   C_{fi\,\lambda}(t;\vec{p},\vec{q})
     =  {}_\lambda\langle 0| \hat{\tilde{B}}_{N_f}(\vec{p})
                            |N(\vec{p})\rangle \times
        {}_\lambda\langle N(\vec{p}) | \hat{\bar{B}}_{N_i}(\vec{0})
                             |0\rangle_\lambda \times
          e^{-E_{N\,\lambda}(\vec{p},\vec{q})t} + \ldots \,,
\label{C_fact}
\end{eqnarray}
where as this equation suggests we have taken the lowest state 
$|N(\vec{p})\rangle$ to be well separated from other states. 
Furthermore we have defined ${}_\lambda\langle N(\vec{p})|$ as
\begin{eqnarray}
  {}_\lambda\langle N(\vec{p}) |
    =  \langle N(\vec{p}) |
       + \lambda_\alpha\, \sum_{E_Y(\vec{p}_Y) > E_N(\vec{p})}
             { \langle N(\vec{p})|\hat{\tilde{{\cal O}}}_\alpha(\vec{q})
                                 | Y(\vec{p}_Y) \rangle \over
                                   E_Y(\vec{p}_Y) - E_N(\vec{p}) }
                        \, \langle Y(\vec{p}_Y) | + O(\lambda^2) \,.
\end{eqnarray}
(We do not give the $O(\lambda^2)$ term here.)
While the final nucleon operator, $\hat{\tilde{B}}_{N_f}(\vec{p})$, has
a definite momentum and so just picks out one state, the initial
nucleon operator, $\hat{B}_{N_i}(\vec{0})$, being at position 
$\vec{x}=\vec{0}$ contains all momenta and states (indicated here by
the sum over $|X(\vec{p}_X)\rangle$). 
For the matrix elements that appear in the modified energy in 
eq.~(\ref{C_fact}), rather than writing them in terms of the operator 
$\hat{\tilde{{\cal O}}}_\alpha$ we first 
use $\hat{O}(\vec{x}) = e^{-i\hat{\vec{p}}\cdot\vec{x}}\,\hat{O}(\vec{0}) \,
                                  e^{i\hat{\vec{p}}\cdot\vec{x}}$
on the relevant term to give
\begin{eqnarray}
   \langle X(\vec{p}_X)| \hat{\tilde{{\cal O}}}_\alpha(\vec{q}) 
                             | N(\vec{p}) \rangle
     = \langle X(\vec{p}_X)| \hat{O}_\alpha(\vec{0})| N(\vec{p}) \rangle
          \, \delta_{\vec{p}_X,\vec{p}+\vec{q}}
       + \langle X(\vec{p}_X)| \hat{O}_\alpha^\dagger(\vec{0})| N(\vec{p}) \rangle
        \, \delta_{\vec{p}_X,\vec{p}-\vec{q}} \,,
\end{eqnarray}
so matrix elements step up or down in  $\vec{q}$.
As this is also valid for $X = N$ then the $O(\lambda)$ term%
\footnote{Namely 
  $-\lambda_\alpha\langle N(\vec{p})|\hat{\tilde{{\cal O}}}_\alpha(\vec{q})
                                   | N(\vec{p}) \rangle$.}
vanishes ($\vec{q} \not= \vec{0}$). Generalising each $\lambda$ inserts 
another $\hat{\tilde{{\cal O}}}$ into the matrix element, so we need an 
even number of $\lambda$s, i.e.\ odd powers of $\lambda$ vanish.
This gives finally
\begin{eqnarray}
   E_{N\,\lambda}(\vec{p},\vec{q})
      &=& E_N(\vec{p})
          - \sum_{E_X(\vec{p}\pm\vec{q}) > E_N(\vec{p})} \left[
                 { | \langle X(\vec{p}+\vec{q}) 
                        |\lambda_\alpha \hat{O}_\alpha(\vec{0}) 
                        | N(\vec{p}) \rangle |^2 \over
                       E_X(\vec{p}+\vec{q})-E_N(\vec{p}) } \right.
                                                          \label{E_pert}\\
      & & \left. \hspace*{1.50in}
                 + { | \langle X(\vec{p}-\vec{q}) 
                        |(\lambda_\alpha \hat{O}_\alpha(\vec{0}))^\dagger
                        | N(\vec{p}) \rangle |^2 \over
                      E_X(\vec{p}-\vec{q})-E_N(\vec{p}) } \right]
                 + O(\lambda^3) \,.
                                                          \nonumber
\end{eqnarray}
We need $E_N(\vec{p}\pm\vec{q}) > E_N(\vec{p})$ ($X=N$ is the worst case)
giving $-1 < \omega < 1$ with $\omega = 2\vec{p}\cdot\vec{q} / \vec{q}^2$.
This is the usual definition of $\omega$ (with $q_0 = 0$), which is
in the safe unphysical region.

What has all this to do with the Compton Amplitude? We now interpret
this result and relate it to the Compton Amplitude. Considering its
Minkowski (${\cal M}$) definition again, eq.~(\ref{T_mink}), and again 
inserting a complete set of states for $t>0$ and $t<0$ with the 
appropriate $i\epsilon$ prescription
\begin{eqnarray}
   T^{\ind ({\cal M})}_{\mu\nu}(p,q)
      &=&
           \sum_X \left[
                 { \langle X(\vec{p}+\vec{q}) 
                        |\hat{O}_\mu(\vec{0}) 
                        | N(\vec{p}) \rangle^* \,
                   \langle X(\vec{p}+\vec{q}) 
                        |\hat{O}_\nu(\vec{0}) 
                        | N(\vec{p}) \rangle
                  \over
                   E_X(\vec{p}+\vec{q})-E_N(\vec{p})-q^0-i\epsilon } \right.
                                                          \nonumber    \\
      & & \left. \hspace*{0.35in}
                 + { \langle X(\vec{p}-\vec{q}) 
                        |\hat{O}^\dagger_\nu(\vec{0}) 
                        | N(\vec{p}) \rangle^* 
                     \langle X(\vec{p}-\vec{q}) 
                        |\hat{O}^\dagger_\mu(\vec{0}) 
                        | N(\vec{p}) \rangle 
                   \over
                   E_X(\vec{p}-\vec{q})-E_N(\vec{p})+q^0-i\epsilon } \right] \,.
\end{eqnarray}
Comparing with the previous result of eq.~(\ref{E_pert}) if we set
$q^0 = 0$ and choose the $\vec{p}, \vec{q}$ geometry so that
$E_X(\vec{p}\pm\vec{q}) > E_N(\vec{p})$, i.e.\ $-1 < \omega < 1$ then
we can also drop the $i\epsilon$ which gives
\begin{eqnarray}
   E_{N\,\lambda}(\vec{p},\vec{q}) 
       = E_N(\vec{p}) 
           - {\lambda_\alpha^*\lambda_\beta
               \over {}_{\rm rel}\langle N(\vec{p})|N(\vec{p}\rangle_{\rm rel}}
                T^{\ind ({\cal M})}_{\alpha\beta}((E_N(\vec{p}),\vec{p}), 
                                               (0,\vec{q}))
           + O(\lambda^4) \,.
\label{Delta_E_general}
\end{eqnarray}
As $T^{\ind ({\cal M})}_{\alpha\beta}(p,q)^* = T^{\ind ({\cal M})}_{\beta\alpha}(p,q)$
then the real part of Compton amplitude is symmetric (unpolarised case with
$\lambda$ real) while the imaginary part is anti-symmetric (polarised with
$\lambda$ complex).

For the DIS case considered here in eq.~(\ref{T33})
where $\mu = \nu = z$; $p_z = q_z = 0$, giving 
$T_{33}(p,q) = {\cal F}_1(\omega, Q^2)$.
So with $O_\alpha \to J_z$ we have finally
\begin{eqnarray}
   \Delta E_{N\,\lambda}(\vec{p},\vec{q})
      \equiv E_{N\,\lambda}(\vec{p},\vec{q}) - E_N(\vec{p})
      = - {\lambda_z^2 \over 2E_N(\vec{p})} {\cal F}_1(\omega,Q^2)
        + O(\lambda^4) \,,
\label{delta_E}
\end{eqnarray}
writing the relativistic normalisation explicitly.


\section{The Lattice}
\label{the_lattice}


We now briefly describe some lattice details. In the Lagrangian in the
path integral we add the equivalent perturbation
\begin{eqnarray}
   {\cal L}(x) 
      = {\cal L}_0(x) + 2\lambda_z \cos(\vec{q}\cdot\vec{x}) J_z(x) \,,
\end{eqnarray}
where rather than considering the complete electromagnetic current
we take the vector current $J^{(q)}_\mu$ to be either $Z_V\bar{u}\gamma_\mu u$
(where $q \to u$) or $Z_V\bar{d}\gamma_\mu d$ ($q \to d$). $Z_V$ has been
previously determined. We only modify the propagators for the valence 
$u/d$ quarks in $\lambda$. So there are no quark-line disconnected terms 
considered here. To include this would require at least very expensive 
dedicated configuration generation.
 
More specifically we consider $2+1$ quark mass degenerate flavours on
a $N_S^3\times N_T = 32^3\times 64$ lattice with a spacing 
$a \sim 0.074\,\mbox{fm}$. (Technically $\beta = 5.50$, 
$\kappa_l = 0.120900$, $l = u$, $d$ or $s$ giving 
$m_\pi \sim 470\,\mbox{MeV}$ and $m_\pi L \sim 5.4$ where $L = aN_S$.)
For more details of the action and configuration generation see 
\cite{Bietenholz:2011qq}. Apart from $\lambda_z = 0$, we use $4$
values of $\lambda_z$, namely $\pm 0.0125$, $\pm 0.025$. $Q^2$ has
$5$ values in the range between $3$ and $7\,\mbox{GeV}^2$ and we make 
$\sim O(10^4)$ measurements for each $\lambda_z$, $Q^2$ pair
(varying $\vec{p}$ is numerically cheap as it is not part of the
source, and hence not connected with the numerically expensive 
fermion matrix inversion).


\subsection{ Kinematic coverage}

\begin{figure}[htb]
\begin{minipage}{0.45\textwidth}

   \begin{center}
      \includegraphics[width=6.50cm]{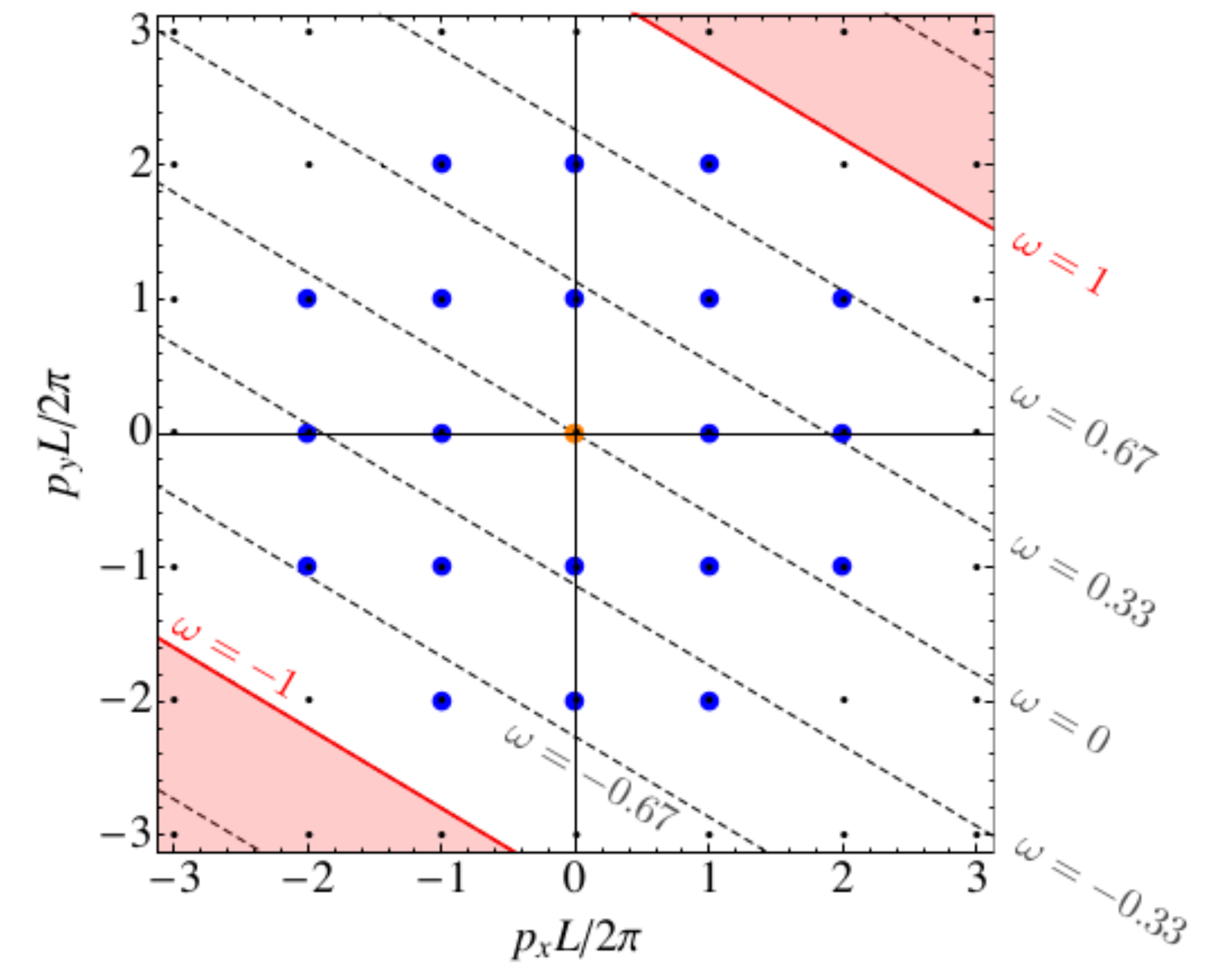}
   \end{center} 

\end{minipage}\hspace*{0.125\textwidth}
\begin{minipage}{0.50\textwidth}

   \begin{center}
      \hspace*{-1.50in}
      \includegraphics[width=8.00cm]{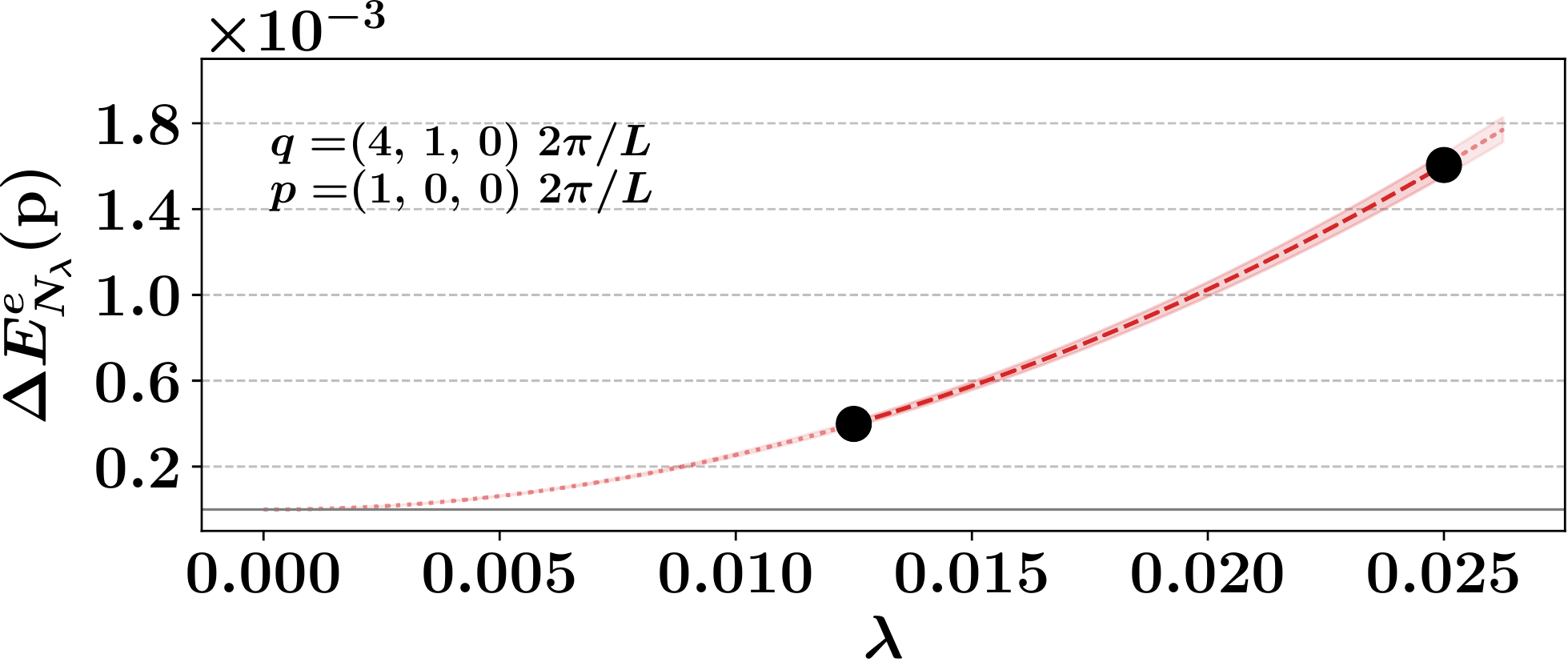}
   \end{center} 

\end{minipage}
\caption{LH panel: The allowed kinematic possibilities for $\vec{p}$
         given $\vec{q} = 2\pi/L(3,5,0)$, where $L =32a$. 
         Lines of constant $\omega = (2 / 34)(3n_x + 5n_y)$ are
         shown dashed. The blue dots give the  allowed momenta.
         RH panel: A plot of $\Delta E_{N\,\lambda}$ against $\lambda_z$
         for $\vec{q} = 2\pi/L(4,1,0)$ and $\vec{p} = 2\pi/L(1,0,0)$.}
\label{kin+eff_energy}
\end{figure}

We now briefly discuss the possible kinematic coverage, which is
sketched in the LH panel of Fig~\ref{kin+eff_energy}.
As an example consider fixed $\vec{q} = (2\pi/L) \,(3,5,0)$.
We can access different $\omega$ by varying the nucleon 
momenta $\vec{p}= (2\pi/L)\vec{n}$ as
$\omega = 2 \vec{p}\cdot\vec{q} / \vec{q}^2 = (2 / 34)(3n_x + 5n_y)$.
Thus for a given constant $\omega$ we have a linear relationship
between $n_y$ and $n_x$ as shown by the lines in the LH panel
of Fig.~\ref{kin+eff_energy}. The blue dots give allowed values of $\vec{p}$.

To extract energy shifts, $\Delta E_{N\,\lambda}$, for each $\lambda_z$ we
form ratios, $R_\lambda$ which isolate the $O(\lambda_z^2)$ term
\begin{eqnarray}
   R_\lambda = {C_{NN\,+\lambda_z}(t)\, C_{NN\,-\lambda_z}(t) \over C_{NN\,0}(t)^2}
           = A_\lambda(\vec{p},\vec{q})\, 
                e^{-2\Delta E_{N\,\lambda}(\vec{p},\vec{q})} + \ldots \,.
\end{eqnarray}
After extracting $\Delta E_{N\,\lambda}$, this is plotted against $\lambda_z$.
An example is shown in the RH panel of Fig.~\ref{kin+eff_energy} for
$\vec{q} = 2\pi/L(4,1,0)$, $\vec{p} = 2\pi/L(1,0,0)$ 
(giving $Q^2 = 4.7\,\mbox{GeV}^2$). A quadratic fit gives from 
eq.~(\ref{delta_E}) the structure function, ${\cal F}_1(\omega,Q^2)$, 
at one value of $\omega$. Repeating this for various values of $\vec{p}$ 
and $\vec{q}$ gives the complete structure function of $\omega$ and $Q^2$.


\subsection{Results}
\label{results}

\begin{figure}[h]

   \begin{center}
      \hspace*{-0.50in}
      \includegraphics[width=9.50cm]
                          {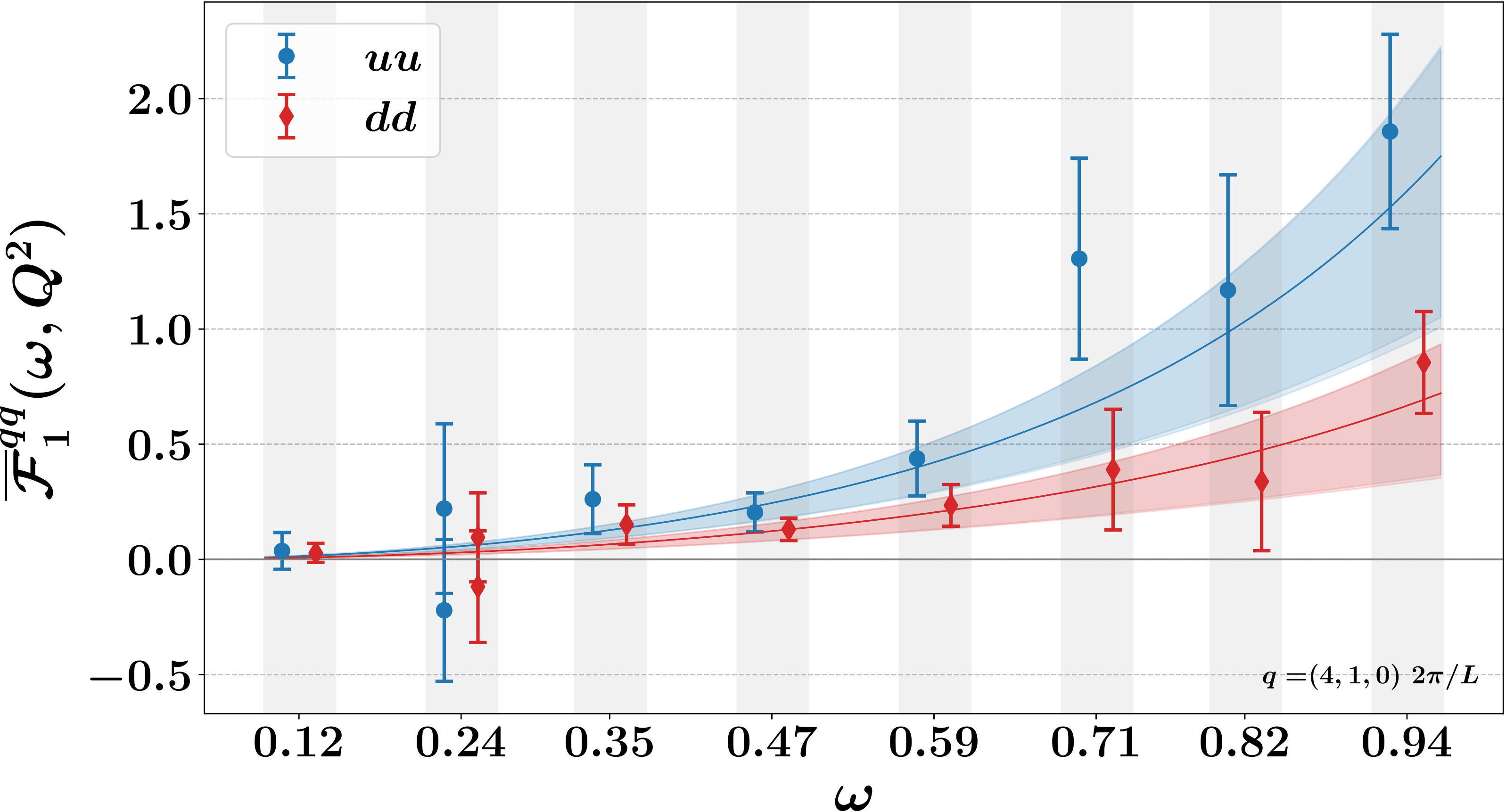}
   \end{center}
\caption{$\omega$ dependence of ${\cal F}_1(\omega,Q^2)$ for 
         $Q^2 = 4.7\,\mbox{GeV}^2$. The blue circles are for 
         $J_z^{(u)}$, the red diamonds for $J_z^{(d)}$. The fits,
         blue and red lines with errors given by shaded region are
         described in the text. The points are slightly shifted 
         for clarity.}
\label{F1_versus_omega}

\end{figure}
\begin{figure}[htb]
\begin{minipage}{0.45\textwidth}

   \begin{center}
      \hspace*{0.25in}
      \includegraphics[width=7.50cm]{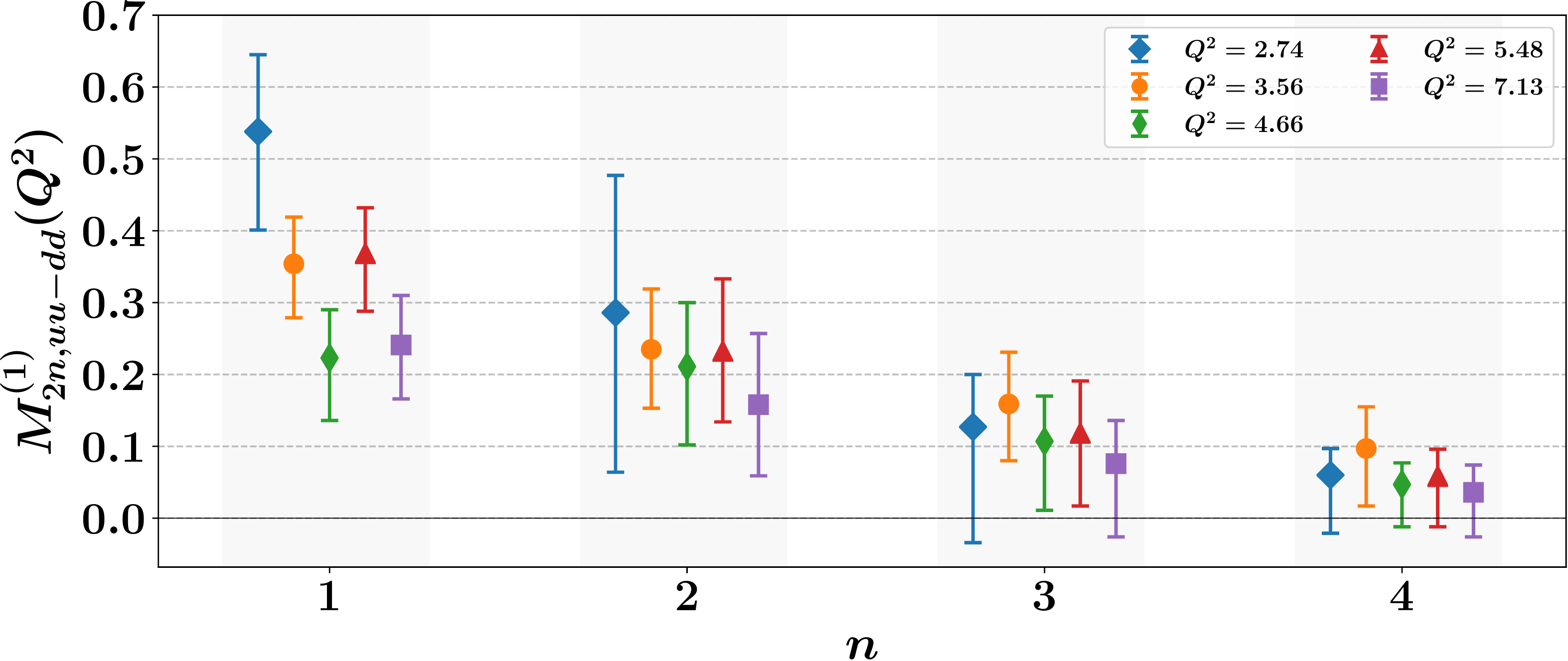}
   \end{center} 

\end{minipage}\hspace*{0.10\textwidth}
\begin{minipage}{0.50\textwidth}

   \begin{center}
      \hspace*{-0.50in}
      \includegraphics[width=5.50cm]{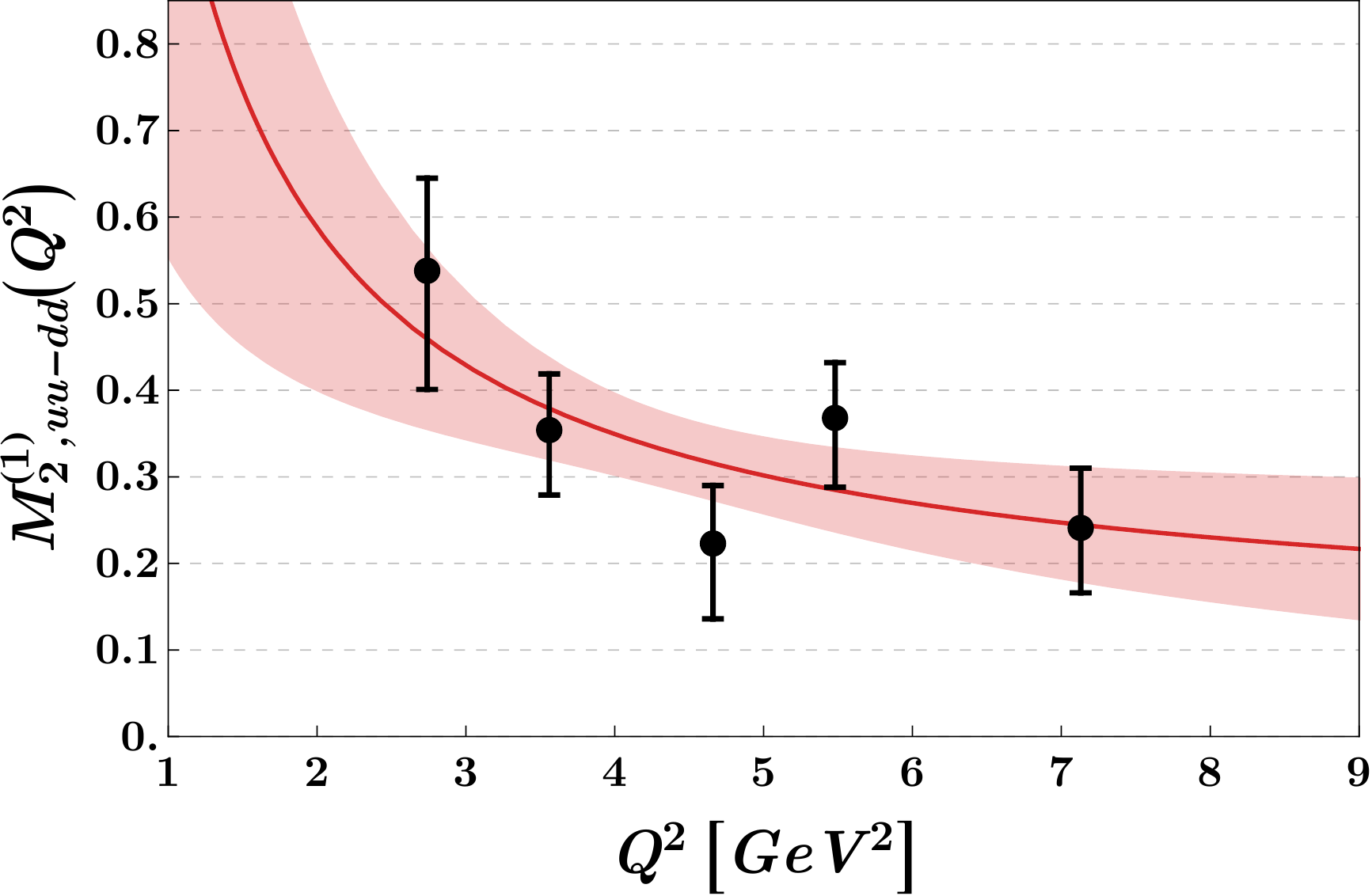}
   \end{center} 

\end{minipage}
\caption{LH panel: The first $5$ isovector moments $M^{(1)}_{2n;u-d}(Q^2)$ 
         (using $J^{(u)}_z -J^{(d)}_z $ for various $Q^2$ values).
         RH panel: The corresponding valence PDF for $M^{(1)}_{2;u-d}(Q^2)$
         $Q^2 = 2.7\,\mbox{GeV}^2$, together with the fit 
         from eq.~(\protect\ref{fit_Q2}).}
\label{moments+power}
\end{figure}

In Fig.~\ref{F1_versus_omega} we show ${\cal F}_1(\omega,Q^2)$ as a 
function of $\omega$ for $Q^2 = 4.7\,\mbox{GeV}^2$ for $J^{(u)}_z$ and 
$J^{(d)}_z$ separately. This figure is our main result. 
We now mention some further consequences from this result. 
From eq.~(\ref{mellin_moments}) we can make a fit  to 
${\cal F}_1(\omega,Q^2)$ to determine the (low) Mellin moments. 
We have the constraints 
$M^{(1)}_{2} \ge M^{(1)}_{4} \ge \ldots \ge M^{(1)}_{2n} \ge \ldots > 0$
for $u$, $d$ separately and so we have implemented a Bayesian procedure
(likelihood with priors as constraints). These are also shown in
the LH panel of Fig.~\ref{moments+power} for $n=6$.
We note that the fall-off of the moments is as expected,
however the second moment does not decrease as rapidly as expected from DIS.

Alternatively we can investigate the $Q^2$ dependence of a particular
moment and investigate scaling and the existence of power corrections 
not restricted to the OPE and large $Q^2$ as shown in the RH panel
of Fig.~\ref{moments+power}. We also made the naive fit
\begin{eqnarray}
   M^{(1)}_{2;u-d}(Q^2) = M^{(1)}_{2;u-d} + {C_2^{(u-d)} \over Q^2} \,.
\label{fit_Q2}
\end{eqnarray}
We concluded, \cite{Can:2020sxc,QCDSF-UKQCD-CSSM:2020tbz,Horsley:2020ltc},
that we need $Q^2 \gsim 16\,\mbox{GeV}^2$ to reliably extract moments 
at a scale of $\mu = 2\,\mbox{GeV}$.

\begin{figure}[h]

   \begin{center}
      \hspace*{-0.50in}
      \includegraphics[width=5.50cm]{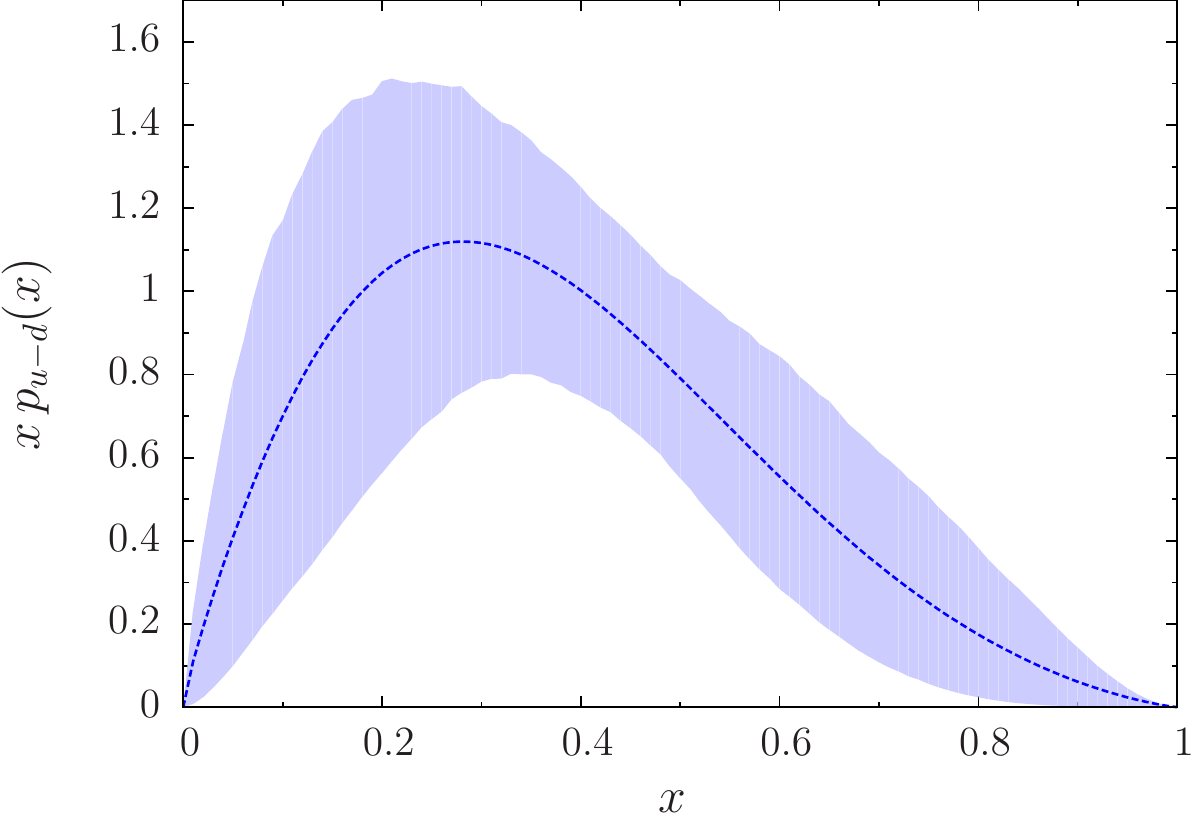}
   \end{center} 

\caption{The valence PDF, $p_{\rm val}^{(u-d)}$ for $Q^2 = 2.7\,\mbox{GeV}^2$.}
\label{reconstruction}
\end{figure}

Is it possible to reconstruct the Form Factor, $F_1$ or indeed the PDF? 
This, of course, would be the ultimate goal. From eq.~(\ref{disp_rel})
we have
\begin{eqnarray}
   T_{33}(\omega,Q^2) 
      = \omega \int_0^1 dx \, K(x\omega)\, F_1(x,Q^2) \,,
   \quad \mbox{where} \quad
   K(\xi) =  4 \, {\xi \over 1 - \xi^2} \,.
\end{eqnarray}
This is a Fredholm integral equation and so an inverse problem, which is 
ill defined. Presently with this data, we have first made the ansatz
\begin{eqnarray}
   F_1(x,Q^2) \equiv a p_{\rm val}(x;b,c)
      =  a \, {\Gamma(b+c+3) \over \Gamma(b+2)\Gamma(c+1) } \, x^b(1-x)^c \,,
\end{eqnarray}
(normalised to $\int_0^1dx \,x p_{\rm val} = 1$). Again with a Bayesian
implementation, we find typical results as in Fig.~\ref{reconstruction}
here for $Q^2 = 2.7\,\mbox{GeV}^2$. The general shape is okay 
(the parton model would give a $\delta$-function at $x = 1/3$, 
which is smeared out by QCD corrections).


\section{Further applications}
\label{further_applications}


Finally we briefly mention some more applications of this method.


\subsection{The $O(\lambda)$ term}


We previously showed that the $O(\lambda)$ terms vanish if 
$\vec{q} \not= \vec{0}$. However for $\vec{q} = \vec{0}$ then it is
possible to determine the baryon charges. For example in \cite{Smail:2021hza}
the tensor charge of octet baryons was determined. 

However we can escape this constraint if there is an degeneracy 
when two (or more) states have the same energy. Then we now have
a matrix of states
$M_{rs} = \langle N(\vec{p}_r)|\hat{\tilde{O}}(\vec{q})
                                     | N(\vec{p}_s) \rangle$
(where $r, s = 1\,,\ldots\,, d_S$, where $d_S$ is the number of degenerate 
states). As before the diagonal elements vanish, but the off-diagonal do not.
This can be diagonalised to give $\Delta E_{N\,\lambda}$. In 
\cite{QCDSF:2017ssq} this was investigated (for $d_S = 2$ in the Breit frame)
and applied to form factors and scattering over a large range of $Q^2$.


\subsection{The $O(\lambda^2)$ term}


While most of our present effort has been directed at the forward
Compton amplitude, we have also started to investigate 
Off-forward Compton Amplitude (OFCA) and GPDs in \cite{Hannaford-Gunn:2021mrl}
where we described the fomalism and determined the two lowest moments.


\subsection{Possible future perspectives}


Possible future perspectives include Spin dependent Structure functions 
and Form factors as indicated in eq.~(\ref{Delta_E_general}),
electromagnetic corrections to the proton -- neutron mass splitting
$M_p - M_n = \delta M^\gamma + \delta M^{m_d-m_u}$ via the
Cottingham formula
\begin{eqnarray}
   \delta M^\gamma = {i \over 2M} { \alpha_{em} \over (2\pi)^2}
                                \int {\eta^{\mu\nu} \over q^2 + i\epsilon}
                                     T_{\mu\nu}(p,q) \,,
\end{eqnarray}
and mixed currents, for example neutrino-nucleon charged weak current
$\nu N \to e X$ or $e N \to \nu X$
\begin{eqnarray}
            W^{\mu\nu} 
               &\equiv& {1 \over 4\pi} \int d^4z \, e^{iq\cdot z} \rho_{ss^\prime} \,
                      {}_{\rm rel}\langle p, s^\prime
                      |[J^\mu_{\rm em}(z), J^\nu_{W,A}(0)]|p, s\rangle_{\rm rel}
                                                        \nonumber  \\
               &=& - i\epsilon^{\mu\nu\alpha\beta} \,
                        {q_\alpha p_\beta \over{2p\cdot q}} F_3(x,Q^2) \,,
\end{eqnarray}
where $J^\nu_{W,A} = \bar{u}\gamma_\nu\gamma_5 d$ the axial part of the weak 
charged current.

A potential problem is including quark-line-disconnected matrix elements.
This needs purpose generated configurations with the fermion determinant 
also containing the $\lambda$ term. For (H)MC for the probability definition
of the action also need a real determinant so fermion matrix must be 
$\gamma_5$-Hermitian which means that
$\lambda^V$ and $\lambda^A$ have to be imaginary (while $\lambda^S$, 
$\lambda^P$ and  $\lambda^T$ are all real). In this case $\Delta E_\lambda$ 
develops an imaginary part. (This is not a problem for the valence sector, 
as this is just an inversion of a matrix.) Simulations are however possible
and this was investigated in \cite{Chambers:2015bka} (at $O(\lambda)$)
for the disconnected contributions to the spin of the nucleon.


\section{Conclusions}


We have described here a new versatile approach for the computation
of matrix elements only involving computation of $2$-point correlation 
functions rather than $3$-pt or $4$-pt which is able to compute 
Compton amplitudes and structure function moments.
Advantages include longer source-sink separations -- so less excited 
states contamination and overcoming fierce operator mixing / renormalisation
issues.


\section*{Acknowledgements}


The numerical configuration generation (using the BQCD lattice QCD 
program~\cite{Haar:2017ubh})) and data analysis (using the Chroma software 
library~\cite{Edwards:2004sx}) was carried out on the 
DiRAC Blue Gene Q and Extreme Scaling (EPCC, Edinburgh, UK) 
and Data Intensive (Cambridge, UK) services, 
the GCS supercomputers JUQUEEN and JUWELS (NIC, J\"ulich, Germany)
and resources provided by HLRN (The North-German Supercomputer Alliance), 
the NCI National Facility in Canberra, Australia (supported by the 
Australian Commonwealth Government) and the Phoenix HPC service 
(University of Adelaide). RH is supported by STFC through grant ST/P000630/1.
HP is supported by DFG Grant No. PE 2792/2-1. PELR is supported in part 
by the STFC under contract ST/G00062X/1. GS is supported by 
DFG Grant No. SCHI 179/8-1. RDY and JMZ are supported by the 
Australian Research Council grant DP190100297.




\end{document}